\documentclass[fleqn,usenatbib]{mnras}

\usepackage{newtxtext,newtxmath}

\usepackage[T1]{fontenc}

\DeclareRobustCommand{\VAN}[3]{#2}
\let\VANthebibliography\thebibliography
\def\thebibliography{\DeclareRobustCommand{\VAN}[3]{##3}\VANthebibliography}

\usepackage{graphicx}	
\usepackage{amsmath}	
 
\usepackage{amssymb}	
\usepackage{pdflscape}	

\title[Evolution of Be-HMXBs as potential ULXs]
{Characteristics and evolution of Be-type high mass X-ray binaries as potential Ultraluminous X-ray Sources}

\author[S. Karino]{
Shigeyuki Karino,$^{1}$\thanks{E-mail: karino@ip.kyusan-u.ac.jp}
\\
$^{1}$Faculty of Science and Engineering, Kyushu Sangyo University, 2-3-1 Matsukadai, 
Fukuoka 813-8503, Japan
}

\date{Accepted XXX. Received YYY; in original form ZZZ}

\pubyear{20XX}

\begin{document}
\label{firstpage}
\pagerange{\pageref{firstpage}--\pageref{lastpage}}
\maketitle

\begin{abstract}

Some ultraluminous X-ray sources (ULXs) exhibit X-ray pulses, and their central sources are thought to be neutron stars.
It has also been suggested that some are transient sources with Be-type donors.
In this study, we use the mass accretion model of a Be-type high mass X-ray binary (BeHMXB) to estimate the conditions under which a giant X-ray burst caused by a BeHMXB exceeds the Eddington luminosity.
Moreover, we investigate the duration for which BeHMXBs can be observed as transient ULXs with bursts above the Eddington luminosity during binary evolutions.
The results indicate that BeHMXBs could be ULXs for a typical duration of approximately 1 Myr.
Comparisons with nearby observed BeHMXBs indicate that many binary systems have the potential to become ULXs during their evolution.
Particularly, a BeHMXB system tends to become a ULX when the Be donor has a dense deccretion disc aligned with the orbital plane.
Because BeHMXBs are very common objects and a significant number of them can become ULXs, we conclude that a reasonable fraction of the observed ULXs could consist of evolved BeHMXBs.

\end{abstract}




\section{Introduction}

Ultraluminous X-ray sources (ULXs) are point-like objects with X-ray luminosities above the Eddington luminosity of stellar-mass objects.
The origin of ULXs was believed to be either stellar-mass black holes with mass accretion above the Eddington limit or intermediate-mass black holes with mass accretion rates below the Eddington limit \citep{kaaret2017}.
Nevertheless, X-ray pulses have been detected associated with the ULX, M82 X-2, indicating that neutron stars could also be a source of ULX radiation \citep{bachetti2014}.
In recent years, several ULXs showing X-ray pulses have been identified and it has been widely recognised that many ULXs likely originate from neutron stars \citep{furst2016, israel2017a, israel2017b, tsygankov2017, carpano2018, doroshenko2018, rodriguez2020}.
Additionally, there is a ULX that is thought to have a neutron star origin because of the presence of a strong magnetic field, although no associated X-ray pulse has been detected \citep{brightman2018}.
Even though there are an increasing number of observations of neutron stars with luminosities above the Eddington limit, the mechanism of the X-ray luminosity and the magnetic field strength of such luminous neutron stars remain under debate \citep{eksi2015, tong2015, mushtukov2015, king2019, erkut2020, mushtukov2021}.

The fraction of ULXs for which neutron stars are the X-ray source is also an interesting issue \citep{fragos2015, shao2015, misra2020, kuranov2020}.
To estimate the fraction of neutron star ULXs, it is important to understand the binary nature of ULXs with accreting neutron stars.
From the current limited number of observations, it appears that neutron star ULXs are a type of high mass X-ray binary (HMXB) with a massive companion \citep{karino2016, king2019}.

HMXBs are traditionally classified as either OB-type or Be-type, depending on the type of donor.
If the system has an OB-type star and a short orbital period, the mass transfer from the donor to the neutron star will be via Roche lobe overflow (RLOF), which provides a large mass transfer rate \citep{bildsten1997, frank2002}.
In this case, mass accretion above the Eddington limit occurs in the neutron star and the X-ray luminosity will exceed the Eddington luminosity if the outflow around the neutron star does not overload it too much \citep{lipunova1999, kosec2018, chashkina2019}.
Note that nearby OB-type HMXBs often accrete via strong stellar wind from the OB donors; nevertheless, the stellar wind accretion is unlikely to exceed the Eddington magnitude \citep{karino2016}.

Conversely, if the donor is a Be-type star, the neutron star accretes a large amount of mass as it passes through the circumstellar disc of the Be star during its orbital motion and emits X-rays for a short duration.
Consequently, Be-type high mass X-ray binaries (BeHMXBs) produce X-ray bursts (Type I bursts) with a period related to their orbital period.
Sometimes, regardless of the orbital period, BeHMXBs also produce giant bursts (Type II bursts) that are brighter than normal bursts \citep{reig2011, rivinius2013}.
Some neutron star ULXs appear to have a Be-type companion; thus, it is an interesting question as to whether BeHMXB systems can achieve X-ray luminosities above the Eddington luminosity \citep{okazaki2013, karino2016, karino2021}.
Table 1 lists the neutron star ULXs that are suggested to be accreted from Be-type donors.

Generally, the X-ray luminosity of BeHMXBs is smaller than that of X-ray binary systems with OB-type donors (the X-ray luminosity in a typical Type I burst is approximately $10^{36} \rm{erg \, s}^{-1}$).
However, BeHMXBs constitute a large fraction of HMXBs \citep{bildsten1997, coe2015}.
Thus, if neutron star ULXs are part of the evolutionary process of BeHMXBs, there may be a significant number of neutron star ULXs (even if no pulses are detected) \citep{inoue2020, king2020, kuranov2020, mushtukov2021}.
The neutron star/black hole fraction in ULXs is not only important concerning the initial mass function of massive stars but is also interesting for estimating the population of gravitational wave sources such as double neutron star binaries.

Given this motivation, to investigate the characteristics of neutron star ULXs, \citet{karino2021} considered two conditions under which a BeHMXB can become a ULX: (1) the mass transfer rate from the Be donor to the neutron star exceeds the Eddington rate and (2) the transferred material accretes to the neutron star at a rate exceeding the Eddington rate.
In this study, we adopt the above conditions to the evolution of binary systems and investigate the conditions under which BeHMXBs can become ULXs. 
Particularly, we consider the length of time that binary systems can be super-Eddington sources in the evolutionary process and investigate their observability.

In the next section, we discuss the conditions under which a BeHMXB can become a ULX.
In Section 3, we introduce an evolutionary model of binary systems.
In Section 4, as a result of numerical calculations, we determine the period during which the X-ray luminosity of a BeHMXB exceeds the Eddington luminosity.
We discuss the characteristics of binary systems that become ULXs on the basis of the conditions presented in Section 2, and we consider their observability.
In Section 5, we compare the observed binary systems with the results of the model calculations.
Additionally, we discuss the limitations of the models and future modelling prospects.
The final section is devoted to our conclusions.

\begin{table*}
	\centering
	\caption{\label{table1}
List of neutron star ultraluminous X-ray sources (ULXs) suggested to have Be-type donors.
For each systems, maximum luminosities, spin periods, orbital periods, donor masses, and orbital eccentricities are shown.	
	}
	\label{tab:Table}
	\begin{tabular}{lcccccl} 
		\hline
		Object & $L_{\rm{X, max}}\ \rm{[erg \, s^{-1}]}$ & $P_{\rm{s}}\ \rm{[s]}$ & $P_{\rm{orb}}\ \rm{[d]}$ & $M_{\rm{d}} \ \rm{[M_{\odot}]}$ & $e$	& Ref. \\
		\hline
		SMC X-3 & $3.5 \times 10^{39}$ & 7.8 & 45 & ${>}3.7$ & $0.23-0.26$ & \citet{weng2017,tsygankov2017} \\		
		Swift J0243.6+6124 & $2.6 \times 10^{39}$ & 9.9 & $27.6/28.3$ &  & $0.1$ & \citet{wilson2018, doroshenko2018}\\
		RXJ0209.6-7427 & $1.6 \times 10^{39}$ & 9.3 & $39/47$ & & & \citet{chandra2020}\\
		\hline
	\end{tabular}
	\\

\end{table*}

\section{Conditions under which a BeHMXB burst exceeds the Eddington luminosity}

In a previous study, \citet{okazaki2013} investigated the conditions under which a BeHMXB causes a giant burst.
The conditions under which the luminosity of such giant bursts exceeds the Eddington luminosity can also be studied to determine the conditions that allow a BeHMXB to become a ULX \citep{karino2021}.

The following has been shown by \citet{okazaki2013} in the case of a massive burst of a neutron star supplied with mass via Bondi–Hoyle–Littleton (BHL)-type accretion from a warped disc around a Be donor.
According to the BHL theory, a neutron star captures material inside the accretion radius,
\begin{equation}
R_{\rm{acc}} = \frac{2 G M_{\rm{NS}}}{v_{\rm{rel}}^{2}},
\label{eq:racc}
\end{equation}
when it interacts with the disc material.
The mass transfer rate from the Be disc to the neutron star is given by
\begin{equation}
\dot{M}_{\rm{T}} = \min \left(1, \frac{R^2_{\rm{RL}}}{R^2_{\rm{acc}}} \right) \times \dot{M}_{0}.
\label{eq:MdotT}
\end{equation}
This equation implies that only material inside the Roche radius,
\begin{equation}
R_{\rm{RL}} = D \frac{0.49 q^{2/3}}{0.69 q^{2/3} + \ln (1 + q^{1/3}) },
\label{eq:Rrl}
\end{equation}
where the gravity of the neutron star is dominant, can be accreted.
Here, $D = a(1-e)$ is the separation of the binary at the periastron passage, where $a$ is the semi-major axis of the orbit, which can be obtained from Kepler's law by specifying the donor mass, $M_{\rm{d}}$, the mass of neutron star, $M_{\rm{NS}}$, the orbital period, $P_{\rm{orb}}$, .and the orbital eccentricity, $e$.
$\dot{M}_{0}$ is the mass transfer rate in the BHL accretion theory given by \citet{hoyle1939}, \citet{bondi1944} and \citet{edgar2004},
\begin{equation}
\dot{M}_{0} = \rho v_{\rm{rel}} \pi R_{\rm{acc}}^{2},
\label{eq:Mdot0}
\end{equation}
where $v_{\rm{rel}}$ is the relative velocity between the orbital motion of the neutron star and the rotational velocity of the Be disc material, which depends on the inclination of the disc surface and the orbital plane of the neutron star.
Here, the inclination angle between the disc and the orbital plane is represented by the angle between the orbital velocity vector and the Keplerian velocity vector of the disc material, $\delta$.
Then, the relative velocity can be written as
\begin{equation}
v_{\rm{rel}}^{2} = v_{\rm{orb}}^{2} + v_{\rm{Kep}}^{2} -2 v_{\rm{orb}} v_{\rm{Kep}} \cos \delta .
\label{eq:vrel}
\end{equation}
Here, $v_{\rm{orb}}$ is the orbital velocity at the periastron point and $v_{\rm{Kep}}$ is the Keplerian velocity of the disc material around the Be star at the periastron passage of the neutron star orbit.
If the Be disc and the orbital plane are aligned, the disc is truncated and becomes smaller.
Thus, a giant outburst may be more likely to occur if the disc is inclined to the orbital plane \citep{okazaki2013}.
Conversely, the inclination angle between the disc and the orbital plane also affects the mass transfer rate via the relative velocity between the orbital velocity of the neutron star and the rotational motion of the Be disc.
A smaller $\delta$ also implies a smaller relative velocity, which allows a larger mass transport rate as a result of BHL accretion and therefore a larger X-ray luminosity \citep{okazaki2013, karino2021}.
The dependence of the X-ray luminosity on the inclination is determined by a combination of these factors.

The density of the Be disc, $\rho$, is modelled here using the density of the inner edge of the disc, $\rho_{0}$, as follows \citep{rivinius2013}:
\begin{equation}
\rho = \rho_{0} \left(\frac{D}{R_{\rm{d}}} \right)^{-7/2},
\label{eq:rho}
\end{equation}
The density of the disc and the rotational velocity of the disc material vary with distance from the Be star; however, here all the physical quantities of the disc are represented by their values at the position $D$ of the periastron point.

The material transferred by the BHL process forms an accretion disc around the neutron star.
If the accretion disc is a standard disc, the accretion falls off slowly on a viscous timescale and does not produce a bright and sharp X-ray light curve as in Type II BeHMXB bursts.
Thus, the disc must be an advection-dominant, slim disc type that accretes quickly \citep{okazaki2013}.
To achieve this, the magnetic radius, $R_{\rm{mag}}$, of the neutron star must be smaller than the photon trapping radius, $R_{\rm{trap}}$.
This condition can be replaced by the condition that the mass transport rate $\dot{M}_{\rm{T}}$ be larger than a certain limit, $\dot{M}_{\rm{sc}}$:
\begin{equation}
\dot{M}_{\rm{T}} > \dot{M}_{\rm{sc}} = 1.7 \times 10^{19} k^{7/9} \mu_{30}^{4/9}.
\label{eq:Mdotsc}
\end{equation}
Here, the normalised magnetic moment of the neutron star is used:
\begin{equation}
\mu_{30} = \frac{B R_{\rm{NS}}^{3}}{10^{30}\ \rm{G \, \rm{cm}}^3},
\label{eq:mu}
\end{equation}
where $B$ is the surface field strength and $R_{\rm{NS}}$ denotes the radius of the neutron star.
$k$ is a parameter that depends on the accretionary geometry; here, we use $k = 0.5$ for the disc accretion.
The above condition, Eq.~(\ref{eq:Mdotsc}), must be satisfied for a BeHMXB to undergo a giant burst.

When Eq.~(\ref{eq:Mdotsc}) is satisfied, the accreting material transferred inwards through the slim disc is eventually trapped by the magnetic field of the neutron star forming an accretion column near the magnetic pole, which emits strong X-rays \citep{pringle1972}.
When the mass transfer exceeds the Eddington rate, the radiation pressure becomes dominant somewhere in the accretion disc and an outflow occurs \citep{lipunov1982, lipunova1999, poutanen2007}.
The radius at which the radiation pressure becomes dominant, $R_{\rm{sph}}$, can be written as
\begin{equation}
R_{\rm{sph}} = 1.4 \times 10^{6} \frac{\dot{M}_{T}}{\dot{M}_{\rm{Edd}}}\ \rm{cm},
\label{eq:rsph}
\end{equation}
where $\dot{M}_{\rm{Edd}}$ denotes the Eddington mass accretion rate \citep{king2017}.
Inside $R_{\rm{sph}}$, the accretion rate of the neutron star eventually decreases to
\begin{equation}
\dot{M}_{\rm{acc}} = \frac{R_{\rm{mag}}}{R_{\rm{sph}}} \dot{M}_{T}
\label{eq:MdotM}
\end{equation}
because of the loss of material via outflow.
Here, the magnetic radius of the neutron star is
\begin{equation}
R_{\rm{mag}} = \left[ \frac{\mu_{30}^{4} }{ 8 k^{2} G M_{\rm{NS}} \dot{M}_{\rm{acc}}^{2} } \right]^{1/7}.
\label{eq:rmag}
\end{equation}
Using the mass accretion rate, $\dot{M}_{\rm{acc}}$, given above, the luminosity of a giant burst of a BeHMXB can be obtained such that
\begin{equation}
L_{\rm{X}} = \dot{M}_{\rm{acc}} c^2,
\label{eq:LX}
\end{equation}
and, if this luminosity is greater than the Eddington luminosity, the system could be a candidate for a ULX.

Note that, in Eq.~(\ref{eq:MdotM}), because $R_{\rm{sph}} \propto \dot{M}_{0}$, the dependence of the mass accretion rate on the mass transfer rate vanishes and ultimately becomes
\begin{equation}
\dot{M}_{\rm{acc}} \propto B^{4/9}.
\label{eq:MB}
\end{equation}
This means that the luminosity of the neutron star-powered ULX only depends on the strength of the magnetic field of the accreting neutron star \citep{lipunov1982}.

Furthermore, previous studies suggest that a giant burst is more likely to occur when the Be disc, which is inclined to the orbital plane of the binary system, is warped by the tidal action of the neutron star \citep{negueruela2001, reig2007, moritani2011, okazaki2013}.
This is because, when the neutron star passes through the warped region, the mass transport rate increases, leading to a large outburst.

The conditions for the Be disc to warp can be expressed as \citep{karino2021}
\begin{equation}
R_{\rm{warp}} < R_{\rm{tr}}
\label{eq:warpcondition2}
\end{equation}
based on a combination of the typical radius of warp ($R_{\rm{warp}}$) induced by tidal action \citep{martin2011, rivinius2013} and the disc truncation radius ($R_{\rm{tr}}$) \citep{paczynski1977}.
These radii are expressed as
\begin{equation}
R_{\rm{tr}} = 0.5 a (1-e)
\label{eq:rtr}
\end{equation}
and
\begin{equation}
\begin{split}
R_{\rm{warp}} & = 4.91 \times 10^{11} \left(1- e^2 \right)^{3/4} \alpha_2^{1/2} h 
\left(\frac{P_{\rm{orb}}}{1 \rm{d}} \right) \\
& \times
\left(\frac{M_{\rm{d}}}{M_{\odot}} \right)^{1/2}
\left(\frac{M_{\rm{NS}}}{M_{\odot}} \right)^{-1/2}
\left(\frac{M_{\rm{d}} + M_{\rm{NS}}}{M_{\odot}} \right)^{1/2}
\left(\frac{R_{\rm{d}}}{R_{\odot}} \right)^{-1/2}\ \rm{[cm]},
\end{split}
\label{eq:rwarp}
\end{equation}
respectively.
Here, $\alpha_{2}$ is the coefficient of the shear viscosity of the Be disc in the vertical direction and the value $\alpha_{2} = 2.73$ is adopted in \citet{martin2011}.
Conversely, \citet{cheng2014}, who investigated the presence of giant bursts in BeHMXBs depending on the warp conditions of the Be disc, uses $\alpha_{2} = 0.5 - 1$ to explain the bimodality of BeHMXBs on the Corbet diagram.
For large values of $\alpha_{2}$, most systems with long orbital periods and large eccentricities are rejected as ULX candidates \citep{karino2021}.
However, observations of BeHMXBs indicate that giant bursts occur even in systems with relatively large eccentricities.
For example, GROJ1008-57, which is a giant burst source, has a highly distorted orbit with an orbital period of 248 d and an orbital eccentricity of 0.65 \citep{coe2007, reynolds2020}.
To avoid rejecting such a system, $\alpha_{2}$ should be a small value.
Because $\alpha_{2}$ is highly indeterminate, it is given here for reference only; the value $\alpha_{2} = 0.5$ is used as a reference value.

In summary, a neutron star undergoing mass accretion from a Be donor to be a ULX must simultaneously satisfy Eq.~(\ref{eq:Mdotsc}) and
\begin{equation}
\dot{M}_{\rm{acc}} > \dot{M}_{\rm{Edd}}.
\label{eq:cond2}
\end{equation}
Additionally, we consider Eq.~(\ref{eq:warpcondition2}) as the warp condition for reference.

\section{Evolution of BeHMXBs and ULXs}

Here, we investigate how long a binary system could produce super-Eddington bursts during its evolutionary process by varying each BeHMXB parameter.
We fix the neutron star mass, $M_{\rm{NS}}$ and radius, $R_{\rm{NS}}$, at 1.4 solar masses and 10 km, respectively.
The neutron star magnetic field, $B$, is typically approximately $10^{12} \rm{G}$ in most of the systems observed via cyclotron resonant scattering and other methods \citep{christodoulou2017}; nevertheless, there is debate concerning the neutron star magnetic fields in ULXs \citep{eksi2015, tong2015, koliopanos2017, king2019, mushtukov2015, mushtukov2021}.
Here, we consider cases from $10^{11} \rm{G}$ to $10^{13} \rm{G}$.

B-type stars typically have masses close to 10 solar masses.
Here, we consider a Be donor mass, $M_{\rm{d}}$, of $8 - 16$ solar masses at the zero age main sequence stage.
The Be donor has already evolved by some amount at the moment of the supernova of the initial primary star and the birth of the BeHMXB.
Although the evolutionary stage of the donor depends on the mass of the initial primary star, here, we uniformly assume that the donor age is 5 Myr at the time of the birth of the BeHMXB.

Be stars eject their mass via stellar winds and/or deccretion discs.
The Be disc is considered to have a base density between $10^{-11} \rm{g \, cm}^{-3}$ and $10^{-10} \rm{g \, cm}^{-3}$, and the density distribution follows the power law given in Eq.~(\ref{eq:rho}) \citep{rivinius2013}.

The observed orbital periods of BeHMXBs, $P_{\rm{orb}}$, range from ten to several hundreds of days.
Additionally, the orbital eccentricity, $e$, has been observed to range from near circular to highly eccentric \citep{townsend2011}.
In this study, we consider the orbital period and the orbital eccentricity as being the main binary parameters and try various combinations of these parameters over a wide range.
There is little observational information concerning the parameter $\delta$, describing the inclination between the orbital plane and the Be disc plane.
Here, we compare three possible values of $\delta$: $\pi/16$, $\pi/8$ and $\pi/4$.

Under the initial conditions described above, we study the time evolution of a binary system comprising a Be donor and a neutron star.
The calculation of the binary evolution is based on the method of \citet{hurley2002}, with modifications to the code used in \citet{karino2020}.
For the donor Be star, however, we use a model with an age of 5 Myr at the start of the evolutionary calculation (using Eqs.(1)-(30) in \citet{hurley2000}).
The donor is assumed to be experiencing mass loss according to the mass loss rate of \citet{vink2001}, including ejection as a disc and loss via stellar wind.
Because of the short lifetime of binary systems, however, this mass loss rate does not have a significant effect on the subsequent evolution.
The binary period and the orbital eccentricity evolve as a result of the mass loss, mass transfer, gravitational wave emission and tidal action; however, they do not change significantly during the short lifetimes of HMXBs because of their large separation ($P_{\rm{orb}} {>}$ several days) \citep{reig2011}.
The base density of the disc and the inclination between the orbital plane and the disc plane are assumed to remain unchanged.
Under these conditions, the evolutionary effects are exclusively due to the evolution of the donor radius.

At the orbital periastron, if the Roche radius of the Be star is less than the radius of the Be star itself, mass transfer via RLOF from the donor occurs instead of BHL accretion from the disc.
In this case, a massive mass transfer may occur and the orbit may rapidly approach a circular orbit or, in some cases, a common envelope may be formed \citep{ivanova2013}.
When the binary mass ratio is large and the donor spin angular momentum is dominant, the binary separation shrinks rapidly due to tidal instability \citep{darwin1879,counselman1973,hut1980,taam2010,eggleton2006}.
Particularly, if the RLOF occurs after the donor has significantly evolved, it is unlikely to be observable as a stationary ULX.
In this study, we focus on the evolutionary stage when the BeHMXB becomes a (transient) ULX; accordingly, we terminate the calculation when the donor fills its Roche lobe.
If the donor does not fill its Roche lobe, the calculation is terminated when the donor evolves to the point where it causes a supernova explosion.
We solve for both the binary and donor evolution and, at the same time, consider the ULX conditions to identify the period when the BeHMXB would be considered a ULX.

\section{Results}

Choosing a set of BeHMXB initial parameters ($P_{\rm{orb}}$, $e$, $\rho_{0}$, $\delta$ and $B_{\rm{NS}}$), we can solve for the binary evolution.
At the same time, we consider the conditions under which the luminosity of the bursts exceeds the Eddington luminosity and calculate the length of time during which the BeHMXB system becomes a ULX.
Figure 1 shows colour contours indicating the length of time during which the burst luminosity exceeds the Eddington luminosity during the binary evolution.
In this case, the initial (zero age) donor mass is assumed to be 12 solar masses.
The horizontal axis indicates the initial orbital period of the binary, and the vertical axis indicates the initial orbital eccentricity.
The magnetic field of the neutron star, the base density of the Be disc and the inclination angle of the Be disc plane to the orbital plane are fixed at $B = 10^{12} \rm{G}$, $\rho_{0} = 5 \times 10^{-11} \rm{g \, cm}^{-3}$ and $\delta = \pi/8$, respectively.
Results are shown for evolutionary calculations of approximately 7500 binary systems with varying initial orbital periods and eccentricities.
The band-like area from the lower-left to the upper right of the figure is the region that satisfies the ULX condition during the binary evolution.
The parameter regimes that do not exceed the Eddington luminosity are shown in black.
Particularly, the black region in the upper left of the plot corresponds to the parameter regime in which the Roche lobe of the donor is filled at the beginning of the binary evolution.
Conversely, the lower right region of the plot mainly corresponds to a parameter regime in which the neutron star orbit does not pass through the high-density region of the Be disc and, therefore, sufficient mass transfer is not possible.

The figure indicates that a binary system with a period of less than approximately 20 d can be observed as a ULX for a certain period during its evolution regardless of the orbital eccentricity.
This result is consistent with the observational indication that Type II BeHMXB bursts are more likely to occur with shorter orbital periods \citep{reig2011}.
The length of time that a BeHMXB system could be a ULX during its evolutionary process depends on the orbital parameters but can be up to 10 Myr and is typically on the order of 1 Myr.
Because the lifetime of a 12 solar mass donor is roughly 18 Myr 
, this time is not short in comparison with the lifetime of the HMXB.
This means that BeHMXBs with short orbital periods or large orbital eccentricities are expected to produce giant bursts that exceed the Eddington luminosity and could be observed as transient ULXs for a significant period of their lifetimes.
The duration of a ULX in the presence of stable mass transfer has been calculated by \citet{kuranov2020} to be from approximately 1 kyr–1 Myr.
Thus, the lifetimes of ULXs with Be donors could be longer than those of ULXs with super-giant donors.

Nevertheless, none of the observed BeHMXBs has binary periods shorter than 10 d.
This may be because such an orbital period is too short for the donor to supply enough mass to the disc or because the tidal action is too strong for the Be disc to form.
Hence, it is also expected that ULXs with Be donors will not have orbital periods of less than 10 d.
In Figure 1, the limit of $P_{\rm{orb}} = 10 \rm{d}$ is shown as a vertical line.
If we consider only the right side of this vertical line as a feasible region, we can see that the possible ULXs with Be donors are limited to systems with relatively large orbital eccentricities.

In the figure, the warp conditions of the Be disc, shown in Eq.~(\ref{eq:warpcondition2}), are indicated by two curves for reference.
This condition varies with the evolution of the donor.
The lower curve corresponds to the condition at the initial stage of the binary, i.e. when the donor age is 5 Myr.
The upper curve corresponds to the parameters when the donor reaches the terminal main sequence stage.
The lower parts of these curves indicate the parameter regime in which the Be disc warps inside the truncation radius.
Note that this condition is subject to the indefiniteness of the viscosity coefficient.
Because the viscosity parameter $\alpha_{2}$ is highly indeterminate, we have adopted $\alpha_{2} = 0.5$ as a likely value for warping
\footnote{
Even with the most optimistic viscous parameter ($\alpha_{2} = 0.5$), a Type II burst source with a long orbital period, i.e. GROJ1008-57, still does not satisfy the warp condition \citep{reynolds2020}.
Although it is necessary to account for the fact that the actual disc is more dynamically warped \cite[for example]{franchini2021}, we will consider the analytical warp condition as a reference.
}.
According to this line, systems with high orbital eccentricities and/or long orbital periods will not produce bright bursts.
Thus, possible ULXs with Be donors are restricted to a limited region in which the orbital period is less than roughly 100 d and the orbital eccentricity is smaller than approximately 0.8.
The observable lifetime of a system with these characteristics as a ULX is from approximately 0.1–10 Myr; however, this depends on the parameters.
Such a lifetime remains sufficiently long compared with the lifetimes of binary systems, such that the observability of objects in these parameter regimes as ULXs is ensured.
Additionally, as we will see later, this region contains many observed BeHMXBs.

\begin{figure}
\includegraphics[width=\columnwidth]{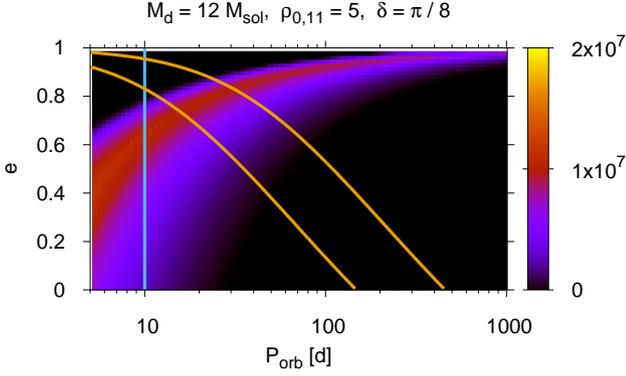}
\caption{
Figure 1: Parameter regime in which a Be-type high mass X-ray binary (BeHMXB) undergoes a burst of super-Eddington luminosity.
The colour contours show the possible duration of the ultraluminous X-ray source (ULX) in years.
The horizontal axis indicates the initial orbital period of the binary, and the vertical axis indicates the initial orbital eccentricity.
The initial mass of the donor is 12 solar masses.
The parameters of the Be disc are fixed to $\rho_{0} = 5 \times 10^{-11} \rm{g \, cm}^{-3}$ and $\delta = \pi / 8$.
The vertical line indicates the lower limit of BeHMXB observations, a period of 10 d.
The two curves show the warp conditions at the beginning of the binary system (upper curve) and at the end of the donor main sequence (lower curve) for reference.
}
\label{fig:result1}
\end{figure}

The maximum possible burst luminosity of a BeHMXB is primarily determined via Eqs.~(\ref{eq:LX}) and (\ref{eq:MB}), which in the case of Figure 1 ($B = 10^{12} \rm{G}$) is approximately $L_{\rm{X}} = 16 L_{\rm{Edd}}$.
The magnetic field affects the maximum X-ray intensity but as only a weak contribution to the ULX condition via Eq.~(\ref{eq:Mdotsc}).
Thus, a variation in the magnetic field of one order of magnitude has little effect on the possible ULX duration.

Conversely, the parameters of the Be disc play an important role.
The density of the disc, $\rho_{0}$, directly affects the mass transfer rate, and the inclination of the disc, $\delta$, affects the mass capture rate in the BHL process.
Namely, a binary system is likely to be observed as a ULX when the disc density is high.
Additionally, a system is likely to be observed as a ULX when the inclination of the disc plane is small.
At the same time, the donor radius changes the size of the disc changes and affects ULX condition accordingly.

Figure 2 shows the possible super-Eddington luminosity bursting duration for different base densities, $\rho_{0}$, of the Be discs.
The other parameters are the same as those in Figure 1.
Figure 2 indicates that the density of the Be disc has a significant effect on ULX formation.
This implies that a Be disc with a low density of less than $10^{-11} \rm{g \, cm}^{-3}$ (see left panel) would not be observed as a ULX because it would not be able to achieve a sufficiently high mass transfer rate to become a ULX.
Conversely, when the base density of the disc increases to $10^{-10} \rm{g \, cm}^{-3}$ (right panel), super-Eddington bursts can be produced even in systems with longer orbital periods.
Based on observations of single Be star discs, Be stars with high-density discs of $5 \times 10^{-11} \rm{g \, cm}^{-3}$ are not rare \citep{touhami2011, rivinius2013, draper2014}.
Thus, it does not seem surprising that even a BeHMXB with typical parameters could become a ULX during its evolution.
When the density of the Be disc is high, the possible ULX duration is extended for systems with large eccentricities.
However, it is not clear whether this is actually observed because the warp condition is more difficult to satisfy for such systems.

\begin{figure*}
\includegraphics[width=6.0cm]{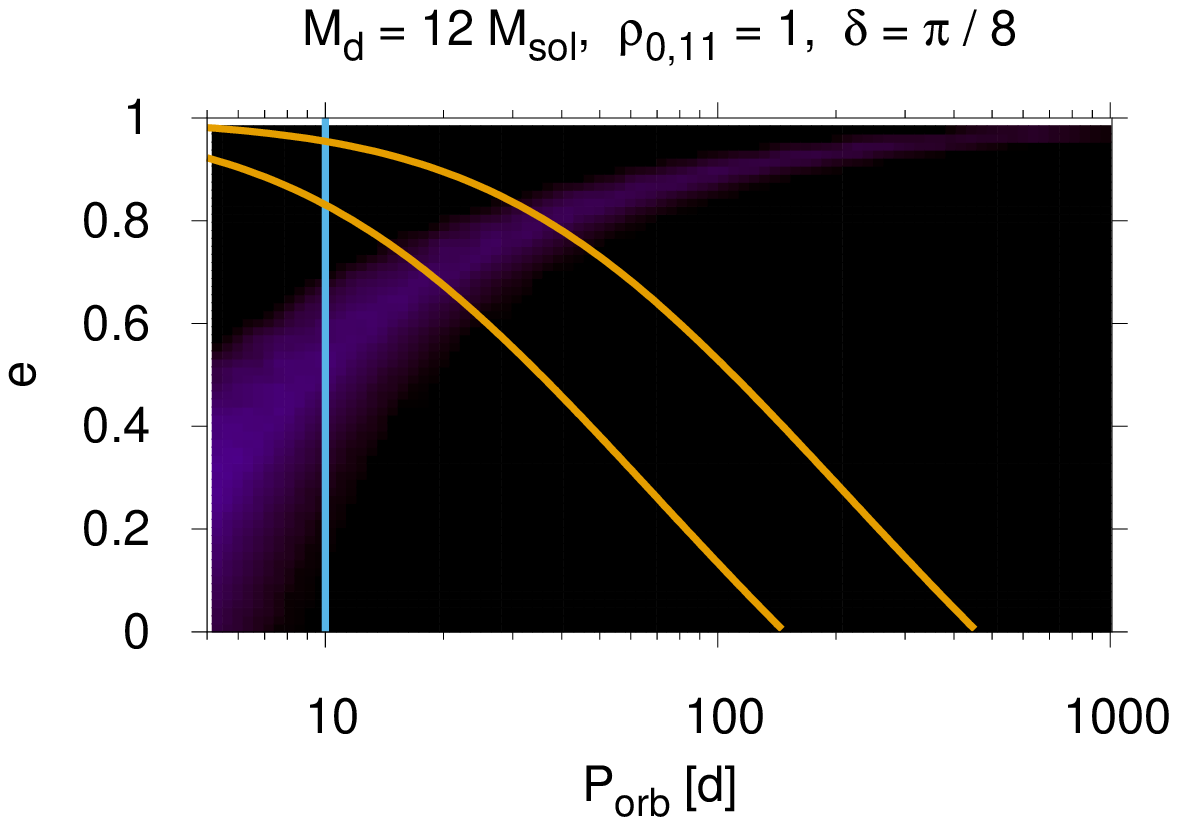}
\includegraphics[width=5.2cm]{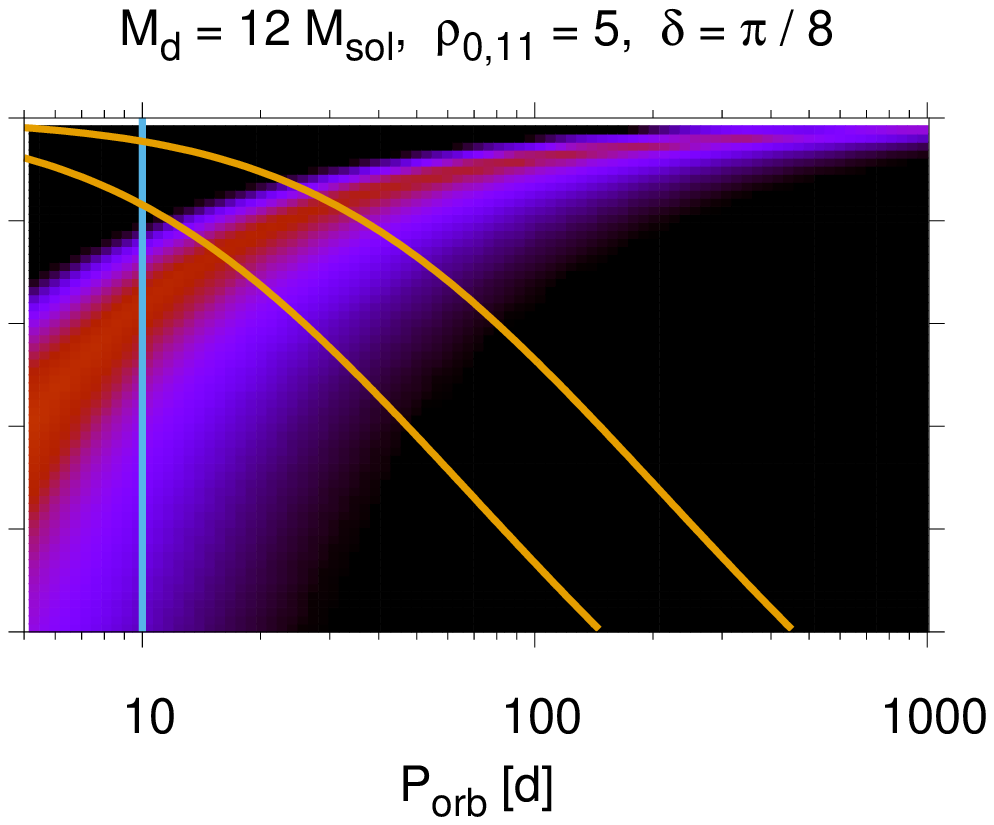}
\includegraphics[width=6.0cm]{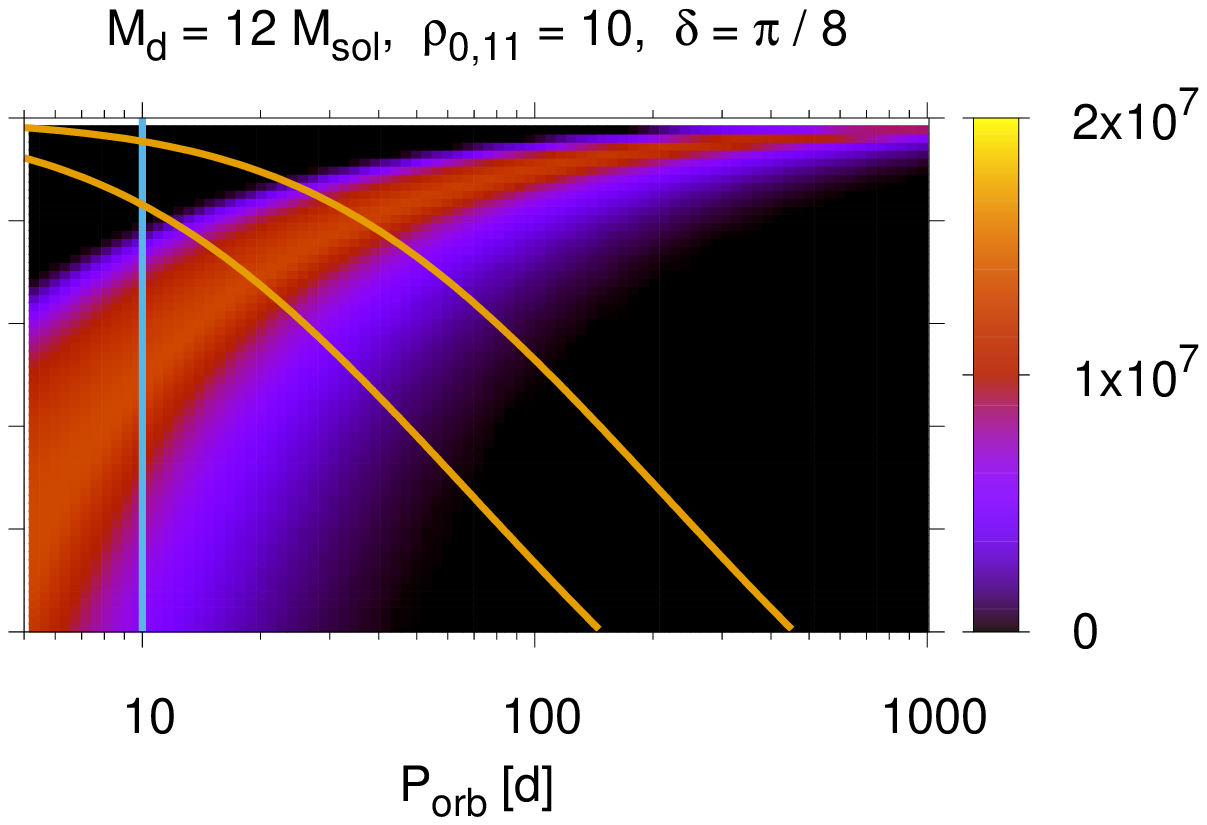}
\caption{
Figure 2: Possible super-Eddington luminosity bursting duration for different base densities of Be discs.
The other parameters are the same as those in Figure 1.
The base densities are, from left to right, $1 \times 10^{-11} \rm{g \, cm}^{-3}$, $5 \times 10^{-11} \rm{g \, cm}^{-3}$ and $10 \times 10^{-11} \rm{g \, cm}^{-3}$.
}
\label{fig:result2}
\end{figure*}

Figure 3 shows the results when the inclination angle $\delta$ is varied from $\pi/16$ to $\pi/4$ to determine the effect of the misalignment between the disc plane and the orbital plane.
The other parameters are the same as those in Figure 1.
The figure indicates that the angle of the orbit also affects the formation of ULXs.
Additionally, the duration of the possible ULX increases when the orbital inclination angle decreases (left panel).
However, this change is primarily observed in the short-period region with a period of less than 10 d and may not be observed in real systems.
Conversely, when the orbital inclination angle is small, the possible ULX region is somewhat wider.
When the disc plane and the orbital plane are close to each other, relatively long-period binary systems with orbital periods of approximately 50 d and not so large orbital eccentricities ($e < 0.5$) can become ULXs for several millions of years.

\begin{figure*}
\includegraphics[width=6.0cm]{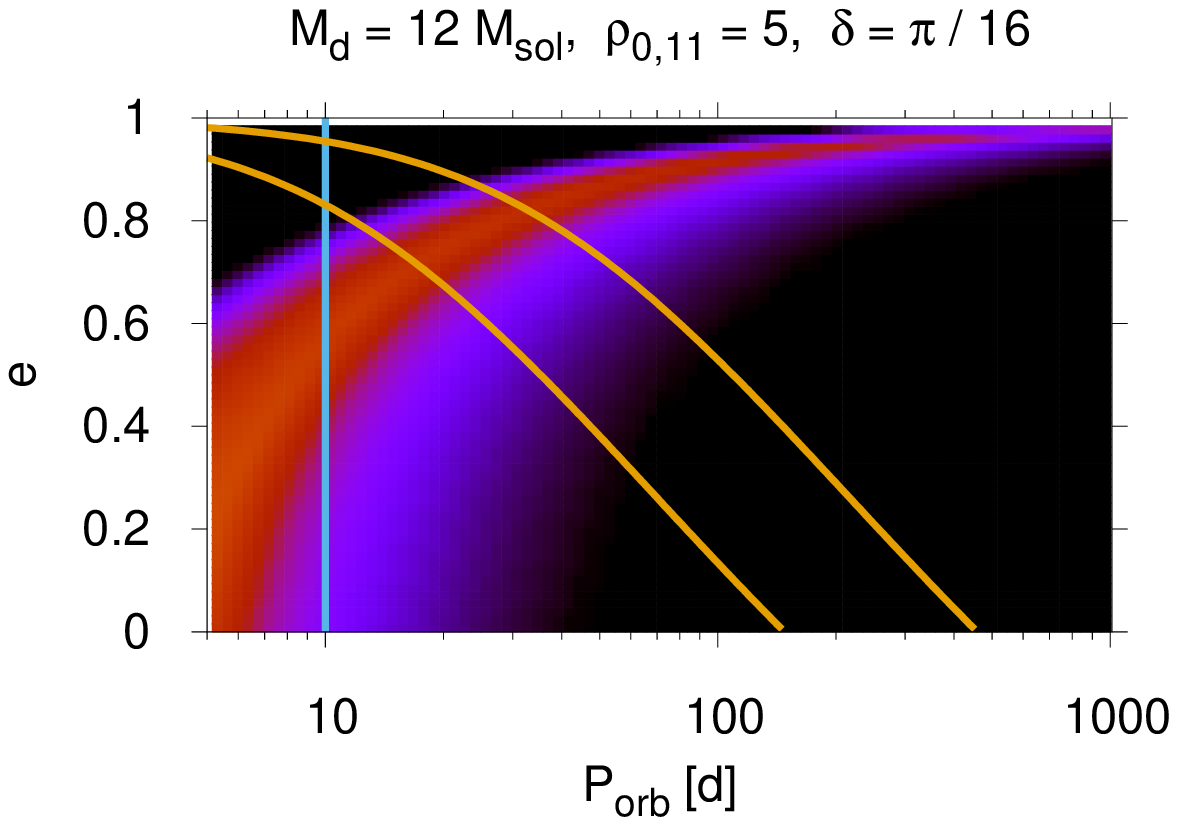}
\includegraphics[width=5.2cm]{modified_m12_b12_r511_d8_warp_line_new.eps}
\includegraphics[width=6.0cm]{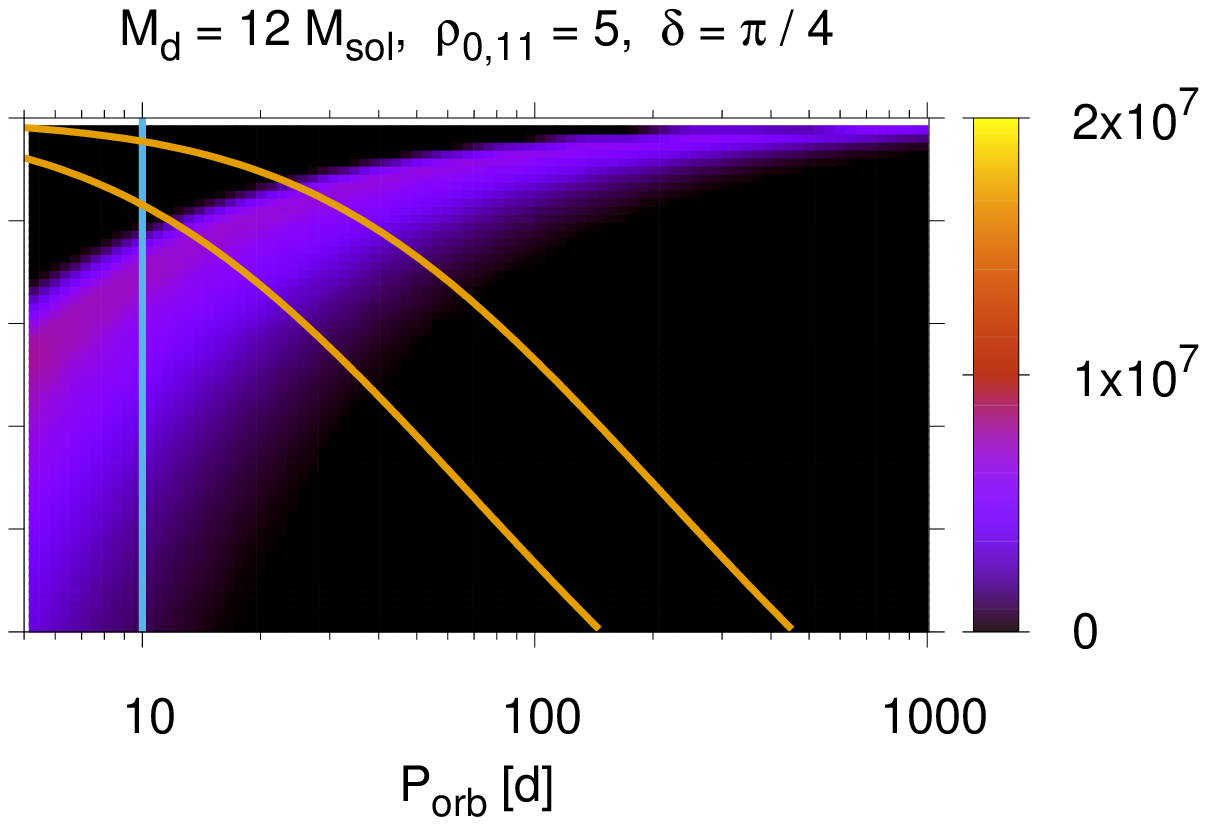}
\caption{
Figure 3: Possible super-Eddington luminosity bursting duration for different inclination angles of the Be disc to the orbital plane.
The other parameters are the same as those in Figure 1.
The inclination angles are, from left to right, $\pi/16$, $\pi/8$ and $\pi/4$.
}
\label{fig:result3}
\end{figure*}

Figure 4 shows the results when the donor mass is varied from $8 M_{\odot}$ to $16 M_{\odot}$.
Again, the other parameters are the same as those in Figure 1.
The difference in the donor mass results in a remarkable difference in the duration of the super-Eddington bursts.
When the donor mass is low (left panel), the lifetime of the ULX increases significantly.
To a certain extent, this result is natural because the lifetime of the donor is significantly extended when its mass is small.
Nevertheless, if the donor mass is large (right panel), the donor radius is also large, such that the dense Be disc extends to the outer edge of the system, which would allow neutron stars to pass through the dense region of the disc even in long-period systems.
According to the computed results, however, this effect is limited.
Instead, the parameter range for giant bursts may become rather narrow because the conditions for the Be disc to warp are more difficult to attain for larger donor masses.

\begin{figure*}
\includegraphics[width=6.0cm]{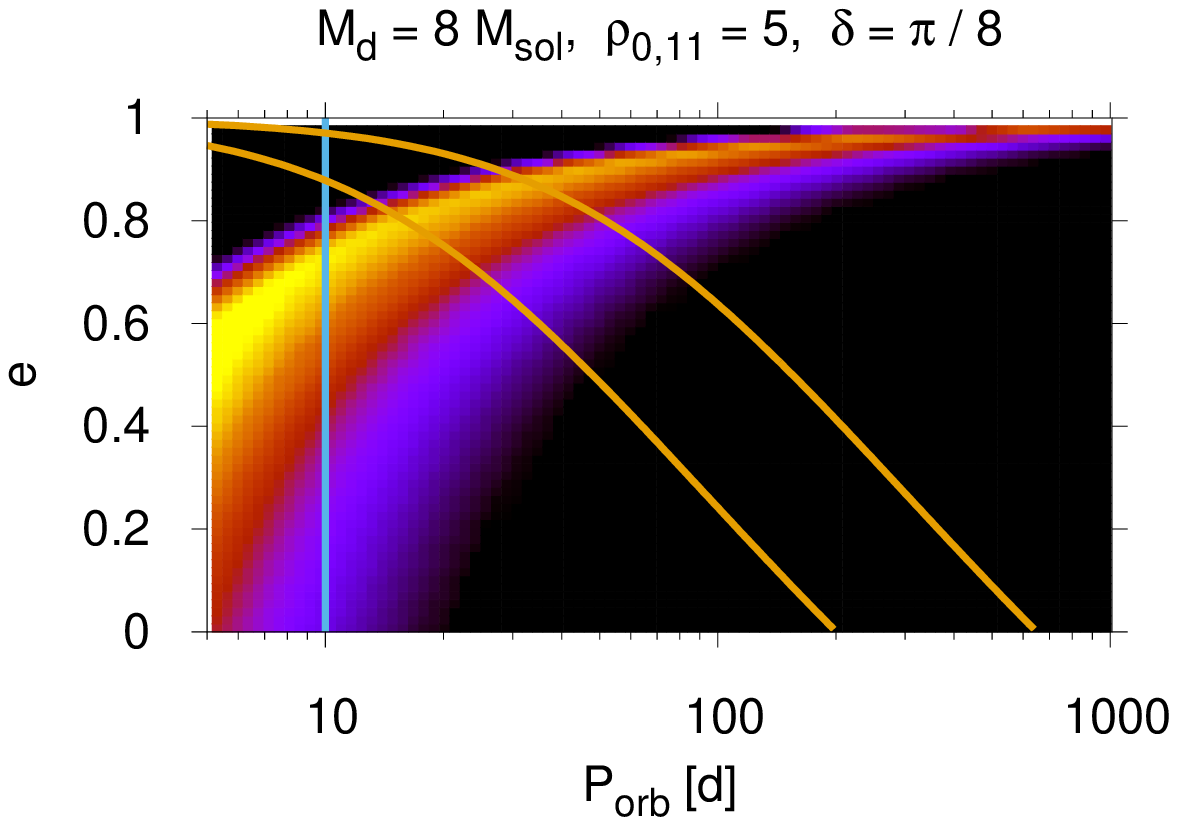}
\includegraphics[width=5.2cm]{modified_m12_b12_r511_d8_warp_line_new.eps}
\includegraphics[width=6.0cm]{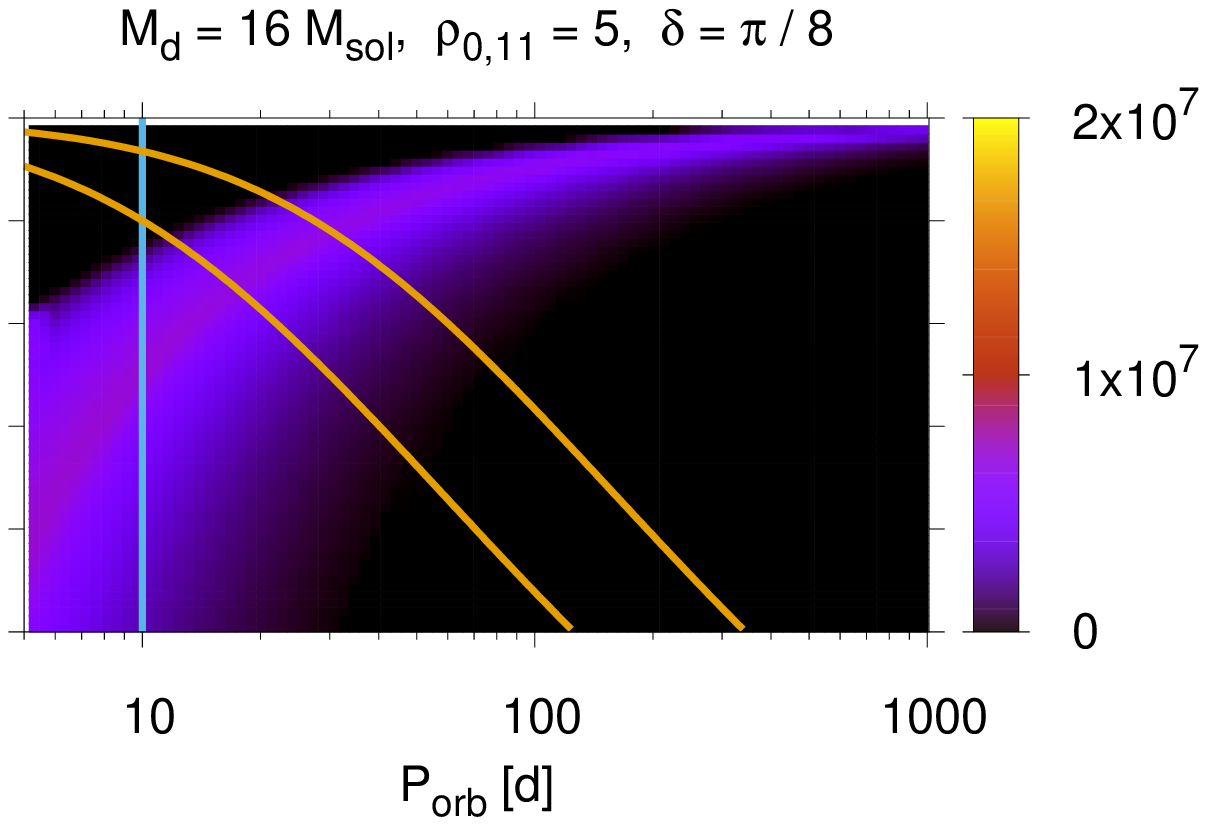}
\caption{
Figure 4: Possible super-Eddington luminosity bursting duration for different donor masses.
The other parameters are the same as those in Figure 1.
The donor masses are, from left to right, 8, 12 and $16 M_{\odot}$.
}
\label{fig:result4}
\end{figure*}

\section{discussion}

As seen in the previous section, BeHMXBs with relatively short orbital periods and large orbital eccentricities can produce X-ray bursts exceeding the Eddington luminosity during their binary evolution.
The duration of these bursts depends on the binary parameters but typically exceeds 1 Myr.
We compare the observed binary parameters of BeHMXBs with the computational results to investigate how many BeHMXBs can potentially evolve into ULXs.
In Figure 5, BeHMXBs observed in our galaxy and in nearby galaxies are plotted as filled circles in the orbital period versus orbital eccentricity plane (data are taken from \citealp{townsend2011} and \citealp{coe2015}).
Two observed ULXs with Be donors whose orbital eccentricities and orbital periods are known are plotted as triangles.
Given the colour contours shown in Figure 1, the edges of the parameter regime that satisfy the conditions for bursts above the Eddington luminosity are indicated by the two violet lines.
The systems in the region between the two lines can produce super-Eddington bursts for durations longer than 1 Myr.
The parameters of the binary are the same as those in Figure 1.
The vertical line shows the minimum orbital period of BeHMXBs, $P_{\rm{orb}} = 10 \, \rm{d}$, for reference.
The curve extending from the upper left to the lower right depicts the warp condition of the Be disc, as in Figure 1, at the terminal main sequence stage for a donor of $12 M_{\odot}$.
The region bounded by these four curves is thought to indicate the parameter regime in which BeHMXBs can be observed as transient ULXs.

If we restrict the sample to systems with known orbital periods and orbital eccentricities, more than half of the BeHMXBs are located in the region of potential ULXs.
Although there may be a bias in that BeHMXBs are more likely to be observed in systems with shorter orbital periods, this suggests that a substantial number of BeHMXBs will generate bursts that exceed the Eddington luminosity during their evolutionary process.
Among the observed ULXs with Be donors, Swift J0243.6+6124 \citep{doroshenko2018, wilson2018}, which has a short orbital period, 
and another ULX, SMC X-3 \citep{tsygankov2017, weng2017}, which has a slightly larger eccentricity, are just outside of the edge of 
the ULX region in Figure 5.
As will be shown later, however, they are included in the ULX region when the parameter is changed slightly.

\begin{figure}
\includegraphics[width=\columnwidth]{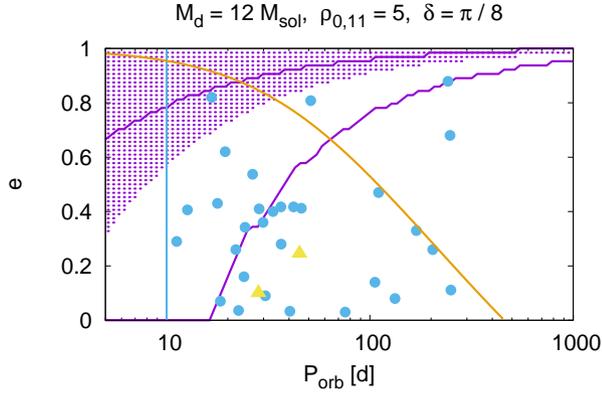}
\caption{
Figure 5: Comparison of observed BeHMXBs with the parameter regimes where super-Eddington bursts can occur.
In the region bounded by the two violet curves, super-Eddington luminosity bursts occur for durations longer than 1 kyr.
The vertical line indicates an orbital period of 10 d, and the orange curve shows the warp conditions at the last stage of the main sequence for reference.
The filled circles indicate the BeHMXBs observed in our galaxy and in the Magellanic Clouds.
The triangles denote ULXs with Be donors (SMC X-3 and swift J0243.6+6124).
The shaded area indicates the region in which Roche lobe overflow starts while the donor is in the main sequence phase.
The binary and donor parameters are the same as those in Figure 1.
}
\label{fig:result5}
\end{figure}

Nevertheless, for a system with a short orbital period or a large orbital eccentricity, the donor could fill its Roche lobe during the evolutionary process.
Before the donor fills the Roche lobe and mass transfer starts, the orbital eccentricity rapidly decreases due to strong tide and the system approaches a circular orbit \citep{hurley2002}.
If the mass transfer is stable, the donor in such a system will lose its circumstellar disc and the system will evolve into a persistent ULX where a large mass transfer rate is achieved via RLOF.
The shaded region in the upper left part of Figure 5 represents the binary parameter regime in which the donor evolves to a larger radius and fills the Roche lobe before it leaves the main sequence.
Figure 5 shows that binary systems that proceed to RLOF during the main-sequence phase are limited to those with short periods or large eccentricities.
Among the observed BeHMXBs, there are few such systems.
Considering that BeHMXBs dominate a large fraction of HMXBs, there may actually be fewer neutron star ULXs accreting by the RLOF.
Rather, neutron stars that show transient bursts following accretion from the Be disc may account for the majority of neutron star ULXs.
Conversely, for systems with orbital periods shorter than 1000 d, RLOF will occur when the donor evolves beyond the main sequence.
Because mass transfer by RLOF beyond the main sequence is expected to be unstable, however, such systems are likely to quickly transition to common envelope evolution \citep{taam2010, ivanova2013}.
In that case, they will radiate brighter at redder wavelengths than at X-ray wavelengths.
The period during which these objects emit brightly is less than a few months.
This period is extremely short compared to the lifetime of the system, so it is rarely observed, although theoretical and observational studies have been carried out in recent years \citep{pastorello2019,blagorodnova2021,matsumoto2022}.

Observational evidence suggests that BeHMXBs with small orbital eccentricities are more likely to produce large bursts \citep{reig2011}.
This may be because there are few systems with large eccentricities, to begin with.
Otherwise, perhaps this is because only systems with small eccentricities can survive as BeHMXBs because systems with large eccentricities quickly enter the RLOF phase.
Because there should be a relationship between the orbital eccentricity and the kick at the time of neutron star formation, the identification of the central object of a ULX may also limit the process of supernova explosions.

As seen in the previous section, the observability as a ULX strongly depends on the parameters of the Be disc.
For example, even a small increase in the density of the Be disc will increase the number of BeHMXB systems that can satisfy the ULX condition at a certain epoch of the evolutionary process.
Figure 6 shows the results when the base density of the Be disc becomes higher: $\rho_{0} = 5 \times 10^{-10} \rm{g \, cm}^{-3}$.
The other parameters are the same as those in Figure 5.
When the disc density increases by one order of magnitude, 
the area of possible ULXs becomes wider and two observed ULXs (Swift J0243.6+6124 and SMC X-3), which were not in this range in Figure 5, are included in the ULX region.
Conversely, as the disc density is reduced, fewer and fewer objects fall within the ULX region.
This suggests that relatively high-density circumstellar discs exist in ULXs with Be donors.

The inclination of the Be disc plane to the orbital plane ($\delta$) also affects the results.
That is, when the vectors of the rotational motion of the Be disc material and the orbital motion of the neutron star are close, the relative velocity of the disc material becomes smaller.
Thus, more material can be captured by the neutron star, the mass transfer rate becomes larger and a super-Eddington burst is more likely to occur.
Consequently, a smaller inclination has the same effect as increasing the density of the Be disc, in the sense that it increases the mass transfer rate.

\begin{figure}
\includegraphics[width=\columnwidth]{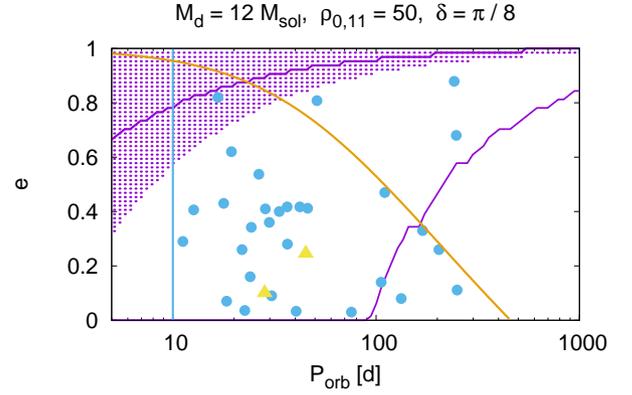}
\caption{
Figure 6: As in Figure 5 but showing the results of increasing the base density of the Be disc by one order of magnitude.
}
\label{fig:result6}
\end{figure}

In the above results, however, the inclination of the disc plane is treated as being constant and independent of the warp condition.
The inclination of the disc plane and the orbital plane could affect the warp conditions.
It is also possible that the rotation axis of the donor and the Be disc plane evolve with time as a result of tidal effects arising from the neutron star.
A dynamic treatment of the warped disc requires numerical simulations and is not considered in depth here (see \citealp{martin2014, franchini2021}).

As well as the inclination of the disc plane, the base density of the disc could evolve as the donor evolves.
Be-type stars are generally thought to spin quickly \citep{rivinius2013}; nevertheless, it is not clear how long the disc is maintained when it spins down as a result of the angular momentum loss associated with the formation of the disc or the tidal forces of the neutron star.
There are many unknowns with respect to the formation of the discs and their densities.
As our understanding of Be-type stars increases, we will be able to learn more about the X-ray emission mechanism of BeHMXBs.

The typical duration for which BeHMXB can produce bursts of super-Eddington luminosity, as calculated here, is approximately 1 Myr.
Nevertheless, the typical lifetime of a BeHMXB is approximately 10 Myr.
Consequently, only approximately 10\% of binary systems in the above parameter regime will actually become ULXs.
Moreover, even if they are in this ULX region, such bursting objects are transient sources, which means that they are not always visible to observers.
Considering the frequency of Type II BeHMXB bursts (less than 10\% of cycles because of the disruption of the Be disc after a large burst \citep{reig2011}), a monitoring period of several years for each object may be necessary to detect a large burst.
Thus, less than approximately 1\% of BeHMXBs are likely to be recognised as active ULXs.

Based on this estimate, there are more than 200 nearby BeHMXBs (of these, 31 objects with known periods and eccentricities are shown in Figure 5).
Conversely, there are three ULX systems with Be donors in the vicinity; SMC X-3, Swift J2043.6+6124 and RX J0209.6-7427 \citep{chandra2020}.
This ratio is not bad compared with the previous estimate.
Because there is an observational bias, further observations are needed.
In principle, it is believed that the observed luminosity is amplified by beaming relative to the intrinsic luminosity \citep{king2017}.
Several recent studies have argued, however, that the effect of beaming is limited \citep{takahashi2017, erkut2020, mushtukov2021}.
The absence of a bias in the viewing angle of the object is convenient for comparisons of the sample with theoretical models.

The potential of a BeHMXB to be a ULX will help in estimates of the number of neutron star binary systems that could be gravitational wave sources and in understanding the ratio of neutron stars to black holes in ULXs \citep{kuranov2021}.
HMXBs containing neutron stars are believed to be the progenitors of double neutron star binary systems.
Particularly, the formation of double neutron star systems, which merge within the age of the universe, requires the passage of moderately spaced HMXBs during their evolution \citep{belczynski2002, tauris2015}.
BeHMXBs, with orbital periods ranging from a few to several hundreds of days, would represent an important fraction of progenitor candidates.
If BeHMXBs can be observed as ULXs, because they can be observed even outside of the Galaxy, their population may provide valuable information for understanding the merger rate of double neutron star binaries.

It is also interesting to determine how many of the ULXs have a black hole origin.
Particularly, the fraction of intermediate-mass black holes is an important question in terms of the formation of massive black holes \citep{kaaret2017}.
If we can classify the ULX sample into those with neutron star origins and those with black hole origins, we will be one step closer to estimating the number of intermediate-mass black holes.
The fact that BeHMXBs, which represent the largest fraction of neutron star binary systems, could be ULXs suggests that the fraction of neutron stars in the ULX sample is not small.
To promote these studies, the model calculations provided in this study can be further developed by adding a method for population synthesis.
Because many results have already been obtained in population syntheses considering ULXs supplied via RLOF \citep{fragos2015, shao2015, wiktorowicz2017, kuranov2020, kuranov2021}, it should be possible to estimate the ULX population more comprehensively by including major mass transport from the Be disc.
These findings are also expected to be useful in extracting information concerning X-ray sources from donor information that will be obtained by next-generation optical telescopes.

\section{Conclusions}

Recently, several ULXs have been discovered that are emitted from neutron stars.
Although the number of observations is still small, some are known to have Be-type donors.
In this paper, we considered the following conditions for BeHMXBs to be ULXs as considered in \citet{karino2021}.
\begin{itemize}
\item The donor does not fill the Roche lobe at the periastron passage.
\item The mass capture rate from the Be disc via the BHL mechanism exceeds the Eddington limit.
\item Even considering the outflow from a slim disc, mass accretion onto the neutron star occurs at a rate exceeding the Eddington limit.
\end{itemize}
Besides this, we considered the following highly indefinite condition.
\begin{itemize}
\item The Be disc warps inside the truncation radius.
\end{itemize}
We hypothesised that BeHMXBs could produce super-Eddington bursts if all these conditions were satisfied, and we searched for the corresponding parameter regime.

The ULX region changes with the evolution of the donor and the binary system.
Particularly, a system with a high-density Be disc aligned with the orbital plane tends to become a ULX.
Additionally, the ULX region depends on the radius of the donor; systems with an expanded donor are likely to produce super-Eddington bursts.
In this study, we also investigated how long a BeHMXB could be a transient ULX with bursts above the Eddington luminosity during the evolution of the donor and binary system.
By investigating the above conditions during the binary evolution, we found that BeHMXBs can typically be ULXs for a duration of approximately 1 Myr during their evolution.
A comparison with nearby BeHMXBs indicates that many binary systems can become ULXs during their evolution.
Because BeHMXBs are common objects and a significant number of them can become ULXs, we conclude that a reasonable fraction of the observed ULXs are BeHMXBs.
Additionally, a significant number of BeHMXBs evolve into stationary ULXs as a result of RLOF caused by the expansion of the donor.
If the ULX population can be deduced from the HMXB population, it can be used to estimate how many non-pulsating ULXs have a neutron star origin.
In the future, it will be necessary to construct a common evolutionary model for Be-type HMXBs and ULXs with Be donors in combination with binary population synthesis computations.

\section*{Data Availability}
 
The data underlying this article will be shared on reasonable request to the corresponding author.

\section*{Acknowledgements}
The author wishes to thank the referee for the helpful suggestions.
This work was supported by JSPS KAKENHI Grant Number 18K03706.



\bibliographystyle{mnras}
\bibliography{ref} 






i


\bsp	
\label{lastpage}
\end{document}